\documentclass{article}

\usepackage{PRIMEarxiv}

\usepackage[utf8]{inputenc} % allow utf-8 input
\usepackage[T1]{fontenc}    % use 8-bit T1 fonts
\usepackage{hyperref}       % hyperlinks
\usepackage{url}            % simple URL typesetting
\usepackage{booktabs}       % professional-quality tables
\usepackage{amsfonts}       % blackboard math symbols
\usepackage{nicefrac}       % compact symbols for 1/2, etc.
\usepackage{microtype}      % microtypography
\usepackage{lipsum}
\usepackage{fancyhdr}       % header
\usepackage{graphicx}       % graphics
\graphicspath{{media/}}     % organize your images and other figures under media/ folder
\usepackage{amsmath}

\title{Modeling of porous lithium metal electrodes: turning the Li-dendrite problem around
}

\author{
  Giovanna Bucci \\
  Palo Alto Research Center Inc. \\
  Palo Alto, CA, 94304 USA\\
  \texttt{gbucci@parc.com} \\
  \AND
  Tushar Swamy \\
  Our Next Energy Inc.\\
  Novi, MI, 29050 USA \\
  \And
  W. Craig Carter \\
  Massachusetts Institute of Technology \\ 
  Cambridge, MA, 02139 USA \\
  \And
  Morad Behandish \\
  Palo Alto Research Center Inc. \\
  Palo Alto, CA, 94304 USA\\
}

\begin{document}
\maketitle

\begin{abstract}
The properties of rechargeable lithium-ion batteries are determined by the electrochemical and kinetic properties of their constituent materials as well as by their underlying microstructure.
Microstructural design can be leveraged to achieve a leap in performance and durability.
Here we investigate a porous electrode structure, as a strategy to increase the surface area, and provide structural stability for Li-metal anodes. 
The porous architecture consists of a mixed electron/ion conductor that function as a scaffold for lithium metal deposition.
A new finite element model was developed to simulate the large topological changes associated with Li plating/stripping. This model is used to predict the current density distribution as a function of material and structural properties.
A dimensionless quantity that combines Li-ion conductivity, surface impedance and average pore size is shown to be a good indicator to predict the peak current density.  
Preventing current localization at the separator reduces the risk of cell shorting. The analyses show that the peak current scales as $(hG)^{1/2}$, where $h$ is the ratio between surface and bulk conductivity and $G$ is the average pore size. 
Stability analyses suggest that the growth is morphologically stable, and that confining Li-plating into pores can enable high-energy density solid-state batteries.
In addition to optimizing porous electrodes design, this finite element method can be extended to studying other Li-battery structures.
\end{abstract}

% keywords can be removed

\section{Introduction}

The need for fast charging, high energy density, long battery life is currently driving the search for a robust solid-state electrolyte which can enable Li metal electrodes. The attempts in using lithium as a negative electrode in Li-ion batteries have revived in recent years, as the energy density of graphite-based cells is reaching its limits, and the demand for high-energy applications is growing.
High-rate, high-capacity cycling of Li metal anodes has been hindered by cell-shorting due to non-uniform Li deposition (forming a dendritic structure), causing the solid electrolyte (SE) layer to fracture~\cite{CHEN202170, CAO202057, KE2020309, KAZYAK20201025, https://doi.org/10.1002/aenm.201701003, Swamy_2018}. To address this problem, we investigate how a porous electrode structure
can self-regulate the local current density and prevent cell shorting.
The porous topology provides a scaffold for Li-deposition and stripping, supporting both mechanical integrity and Li-accessibility.
The goal of this study is to understand what characteristics (geometry and material properties) are required to achieve stable cycling 
%of high-energy-density batteries 
at current densities relevant to vehicle electrification.

To date, several strategies have been developed to mitigate the dendritic growth of metallic Li. In particular, Li metal storage in three-dimensional (3D) framework electrodes is considered an effective dendrite growth mitigation strategy since these electrodes with large surface areas and high porosity result in reduced local Li plating current densities and further can confine metallic Li within their porous architectures resulting in increased tolerance to volume expansion and associated particle fracture effects~\cite{CHEN202170, doi:10.1021/acsenergylett.9b01987, https://doi.org/10.1002/adfm.201606422, doi:10.1021/acsami.1c12576, JIN2017177, doi:10.1021/acs.nanolett.8b01295, https://doi.org/10.1002/adma.201707132, doi:10.1021/acsenergylett.0c01619, https://doi.org/10.1002/adfm.201808468, HYEON202095, https://doi.org/10.1002/adma.201703729, https://doi.org/10.1002/adfm.201905940, doi:10.1021/acsaem.9b02101, ZHAO2020123691, YANG201688, ZHANG2021102167}. Promising porous electrode strategies include 3D CuZn current collectors~\cite{doi:10.1021/acsenergylett.9b01987}, Cu current collector and Li composite anodes~\cite{https://doi.org/10.1002/adfm.201606422}, porous carbon electrodes embedded with high surface area N-doped carbon crystals~\cite{doi:10.1021/acsami.1c12576}, and with lithiophilic ZnO quantum dots~\cite{JIN2017177}. Effective porous electrode strategies for all-solid-state systems include 3D electron-ion conducting frameworks such as porous garnet electrolytes with carbon nanotube coatings~\cite{doi:10.1021/acs.nanolett.8b01295}, and 3D printing enabled porous garnet microstructures~\cite{doi:10.1021/acs.nanolett.8b01295, doi:10.1021/acsaem.9b02101}. Most of these strategies, however, exhibit uneven Li plating-stripping because the ionic resistance of the electrolyte filled pores leads to preferential plating-stripping closer to the separator interface~\cite{doi:10.1021/acsenergylett.0c01619}. To address this issue, recent research has focused on (i) microstructural engineering and (ii) lithiophillic surface treatment, in order to facilitate more uniform plating-stripping across the entire surface area of the porous electrode~\cite{https://doi.org/10.1002/adfm.201808468, HYEON202095, https://doi.org/10.1002/adma.201703729, https://doi.org/10.1002/adfm.201905940, ZHAO2020123691}. 

These examples of mesoscale architecture motivate the 
development of a modeling framework to study the relationship between structure, performance and durability in porous electrodes.
Modeling of energy storage materials has, so far, focused on composite electrodes~\cite{garcia2004microstructural, bucci2017modeling, grazioli2016computational, bucci2017effect, bucci2018mechanical,
bielefeld2018microstructural, prifling2019parametric, finsterbusch2018high, stephenson2011modeling}. 
Here we present a computational method developed to track material growth driven by electro-deposition within a complex geometry.
Employing a phase field method to track the Li interface is a possible approach~\cite{wheeler1993computation, karma1999phase, suzuki2002phase, steinbach2009phase, hong2018phase, zhang2021electrochemical, wang2020application, gao2020phase, chen2015modulation, chen2021phase}.
However, we expect Li-metal regions to not simply merge, but to come in contact, deform and exert stress on the electrode walls.
For this reason, we choose to implement a finite element method with a sharp interface representation, and we anticipate adding stress and fracture in future work. 
We also anticipate that as pores begin to fill, there will be additional terms in the Li diffusion potential. Additionally,
the volume change as lithium comes out of solution and plates within the pores creates difficulties in the phase field method.

\section{Methods}

\begin{figure}
     \centering
     \includegraphics[width=\textwidth]{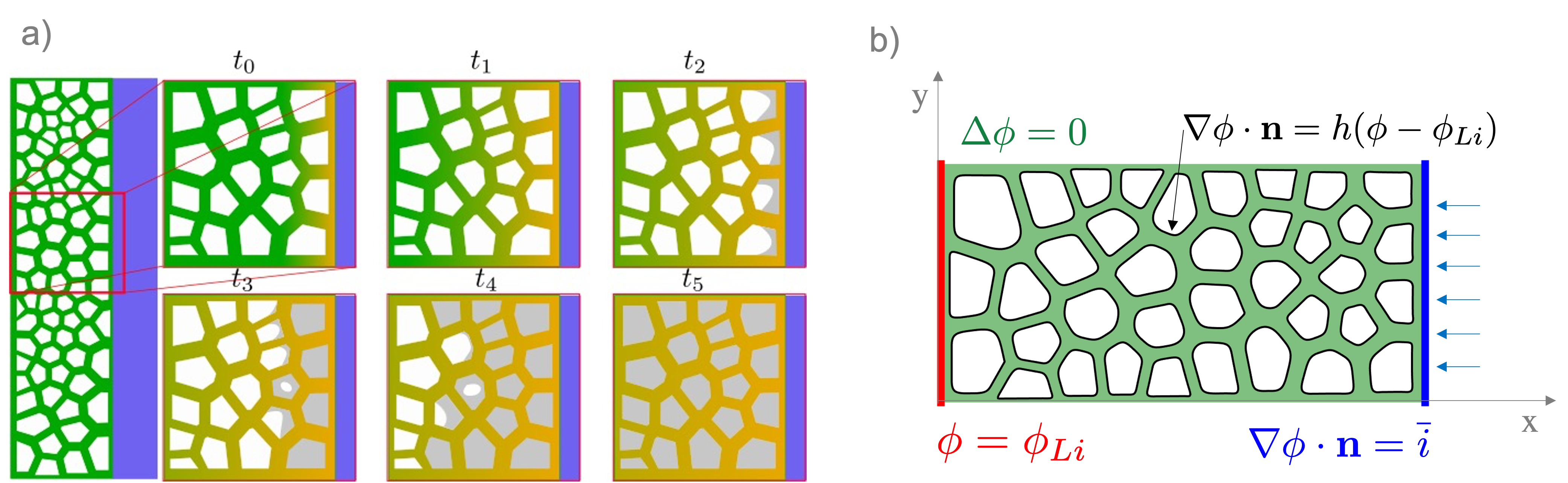}
        \caption{a) Cartoon representation of the anode porous architecture consisting of a mixed electron/ion conductor that function as a scaffold for lithium metal deposition. At discharge, Li progressively fills the pores close to the separator (blue region on the right), and over time the deposition front moves towards the current collector.
        b) A representative section of the lattice with a Voronoi structure is modeled with a finite element approach. The first part of the algorithm calculates the deposition rate at the inner boundary (marked in black). A fixed current is applied at the separator interface (blue line) and fixed voltage is imposed at the interface with current-collector (red line on the left). Zero-flux boundary condition is imposed at the top and bottom edge of the domain.}
        \label{fig:1}
\end{figure}

A finite element model was implemented to simulate electro-deposition in a porous electrode structure. 
In the simulation, the deposition front moves as the cells is the microstructure are filled with lithium (Fig.~\ref{fig:1}).
We assume that the Li plates at the Li/Vapor interface and not the Li/electrolyte interface.
The cathode overpotential is the driving force for lithium deposition and it is calculated by solving the electrochemical problem described below. We consider the domain sketched on the left of Fig.~\ref{fig:1} as representing a porous solid mixed electron/ion conducting material. Steady-state conduction of Li-ion across the electrode is modeled by the Laplace equation as in Eq.~\ref{eq:GaussLaw}, with the potential of lithium $\phi (x,y) $ treated as independent variable. 
A fixed current density is imposed at the separator interface $\partial \mathcal{P}_{+}$ (blue right edge in the figure in Fig.~\ref{fig:1}). 
In Eq.~\ref{eq:neuman}, the constant $\overline{i}$ is the ratio between the current density and the electrolyte bulk conductivity $\kappa$. 
The kinetics at the inner Li/Vapor interface is described by a linear approximation of the Butler-Volmer equation, with the convective coefficient $h$ representing the ratio between surface and bulk conductivity. This is the Robin boundary condition in Eq.~\ref{eq:robin} and applies to the charge-transfer interface.
At the negative electrode, the potential is held constant at $\phi_{\text{Li}}$ by the Dirichlet boundary condition Eq.~\ref{eq:dirichlet}. A zero-flux boundary condition is assumed for the top and bottom edge of the domain.

\begin{subequations}
\begin{align}
    & \Delta \phi = 0 \qquad  \text{on } \mathcal{P} \label{eq:GaussLaw} \\
    & \nabla \phi \cdot \mathbf{n} = h (\phi - \phi_{Li}) \qquad  \text{on } \partial \mathcal{P}_{\text{in}} \label{eq:robin} \\
    & \nabla \phi \cdot \mathbf{n} = \overline{i} \qquad  \text{on } \partial \mathcal{P}_{+} \label{eq:neuman} \\
    & \phi = \phi_{Li} \qquad \text{on } \partial \mathcal{P}_{-} \label{eq:dirichlet}
\end{align}
\end{subequations}

A two-dimensional model representation is chosen because it provides phenomenology that will translate to 3D. 
For a square lattice, in the limit of an infinitely thin wall, the surface area $S$ scales linearly with the number of cells in each dimension as $S = 6 n_x n_y n_z$.
The current density distribution changes with the distance from the separator (x axis in Fig.~\ref{fig:1}), and it is on-average uniform in the other two directions.

For this study we consider microstructures derived from a centroidal Voronoi tessellation, as illustrated in Fig.~\ref{fig:1}. The pore structure will generally depend  on the fabrication method. Here we choose an isotropic structure with closed cells, to demonstrate the model capabilities and derive initial scaling laws.

Electro-deposition is modeled as surface accretion, by assuming that material growth is normal to the Li/Vapor interface.
The numerical implementation of the model was coded in the Wolfram Language and simulated using Mathematica~\footnote{ https://www.wolfram.com/mathematica. Notable functions used from Version 13.0.0 of Mathematica: NDSolve, BSplineFunction, BoundayMesh, ToElementMesh, BoundaryMeshRegion}. Some of the key features of the algorithm are listed below.

\subsection{Algorithm}
\begin{enumerate}
    \item The first step involves solving Eq.~\ref{eq:GaussLaw} to compute the current density $i$ along the charge-transfer interface $\mathcal{P}_{\text{in}}$. The current density determine the amount of material plated at the interface. The normal velocity field is computed as $v_{in} = \Omega_{Li} (i/F)$, with $F$ being the Faraday constant and $\Omega_{Li}$ the partial molar volume of Li. The left image below shows electric potential isolines; the right image illustrate the velocity field at the Li/Vapor interface.
    \begin{minipage}{\linewidth}
    \vspace{5mm}
    
            \includegraphics[width=0.3\textwidth]{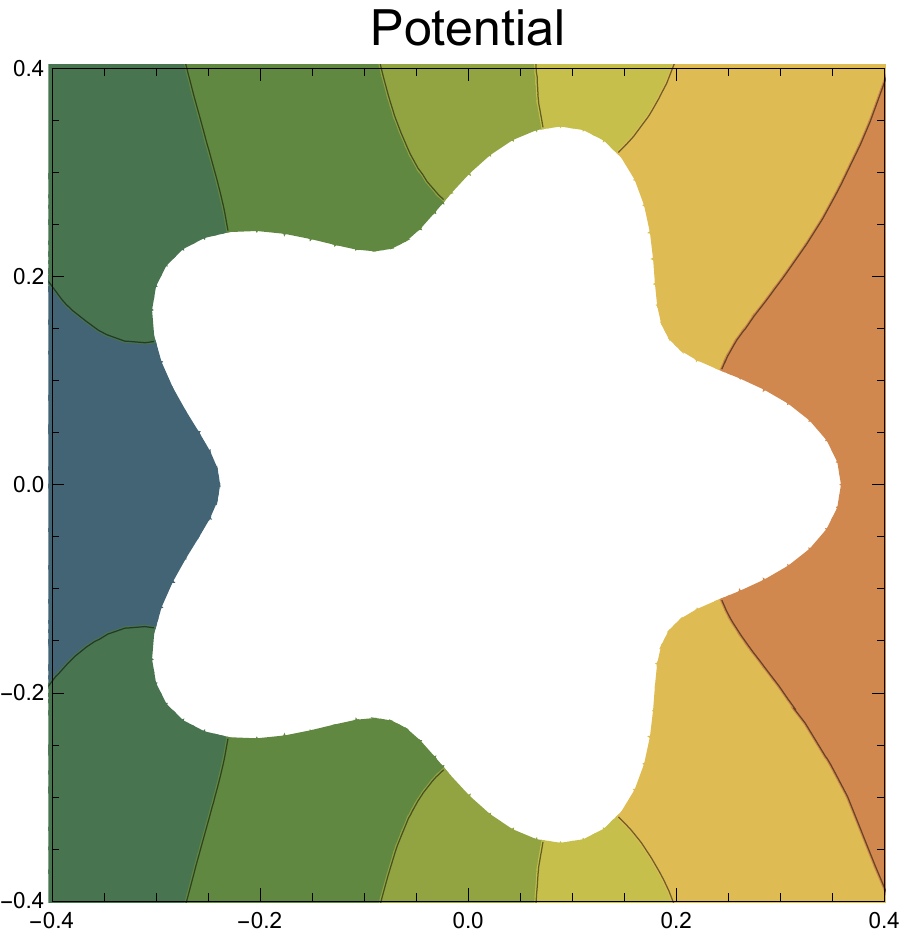}
            \hspace{3mm}
            \includegraphics[width=0.3\textwidth]{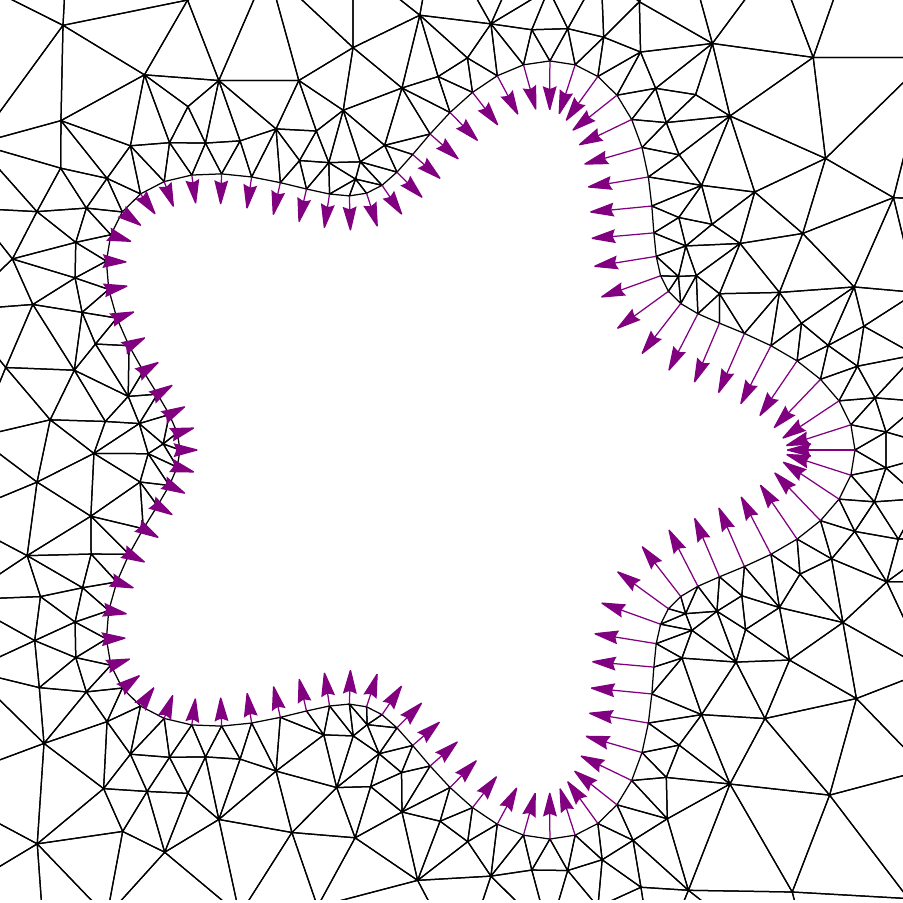}
            \hspace{3mm}
       
    \vspace{5mm}
            % \captionof{figure}{cc}
        \end{minipage}
    
    \item The product between the normal velocity and the time increment $\Delta t$ determines the new location of the boundary nodes (illustrated in pink on the left image below). Some regions show high density of points, corresponding to finite elements with high aspect ratio. Re-sampling of the boundary nodes with uniform spacing is performed at each time step to prevent element inversion, and maintain solution accuracy.
    The re-meshing algorithms is robust with respect  convex/concave geometries and non-uniform current distribution.
    \begin{minipage}{\linewidth}
    \vspace{5mm}
    
            \includegraphics[width=0.3\textwidth]{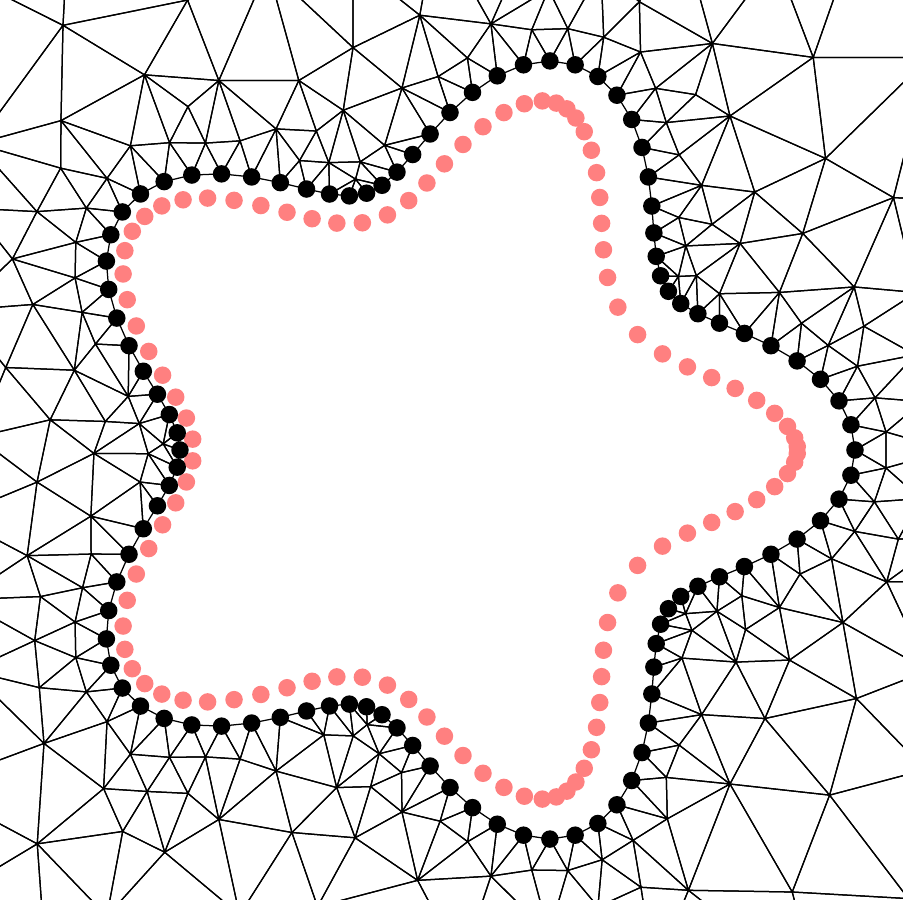}
            \hspace{3mm}
            \includegraphics[width=0.3\textwidth]{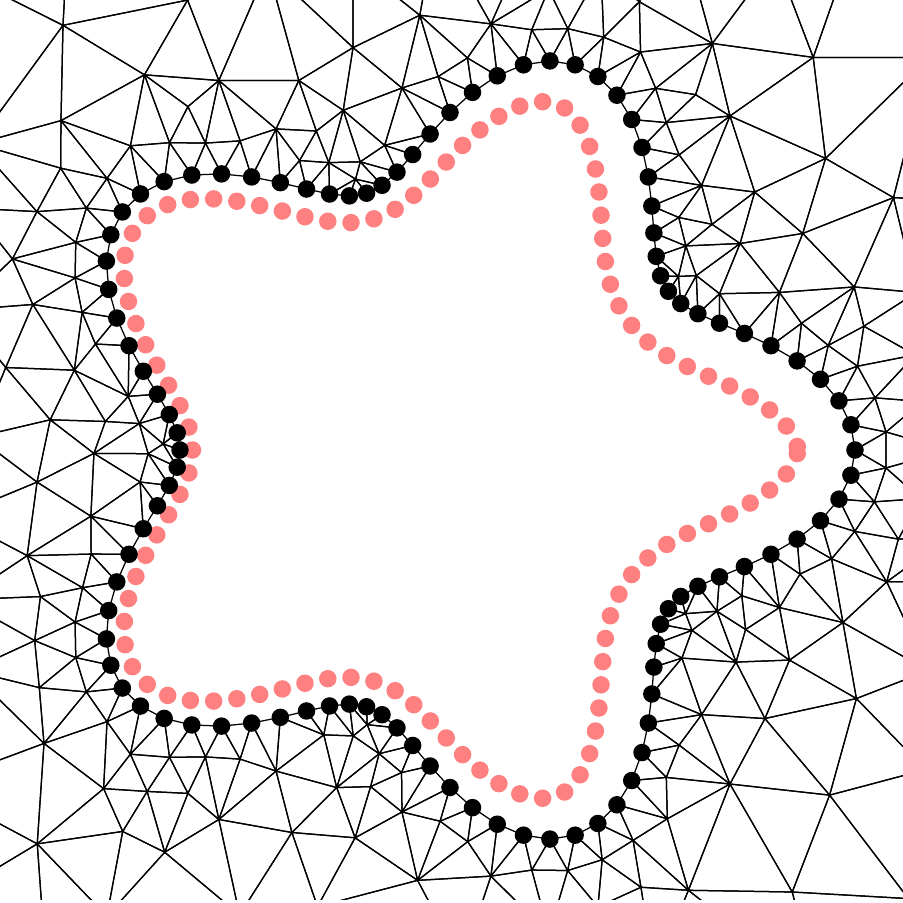}
            \hspace{3mm}
            \includegraphics[width=0.3\textwidth]{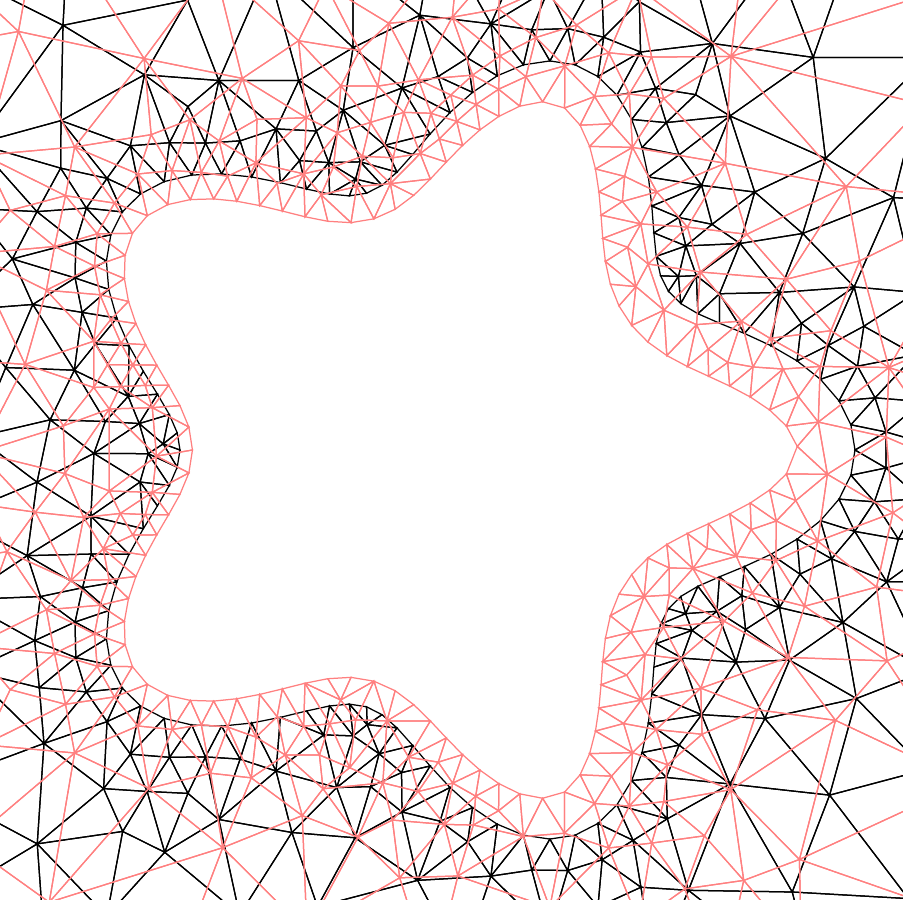}
            
    \vspace{5mm}
            % \captionof{figure}{cc}
        \end{minipage}
    
    \item Acute angles or large time steps may cause self-intersection of the boundary. An algorithm was devised the check for self-intersection and remove the nodes within the intersecting segment as shown in the two images below
    
    \begin{minipage}{\linewidth}
       \vspace{5mm}
       
            \includegraphics[width=0.3\textwidth]{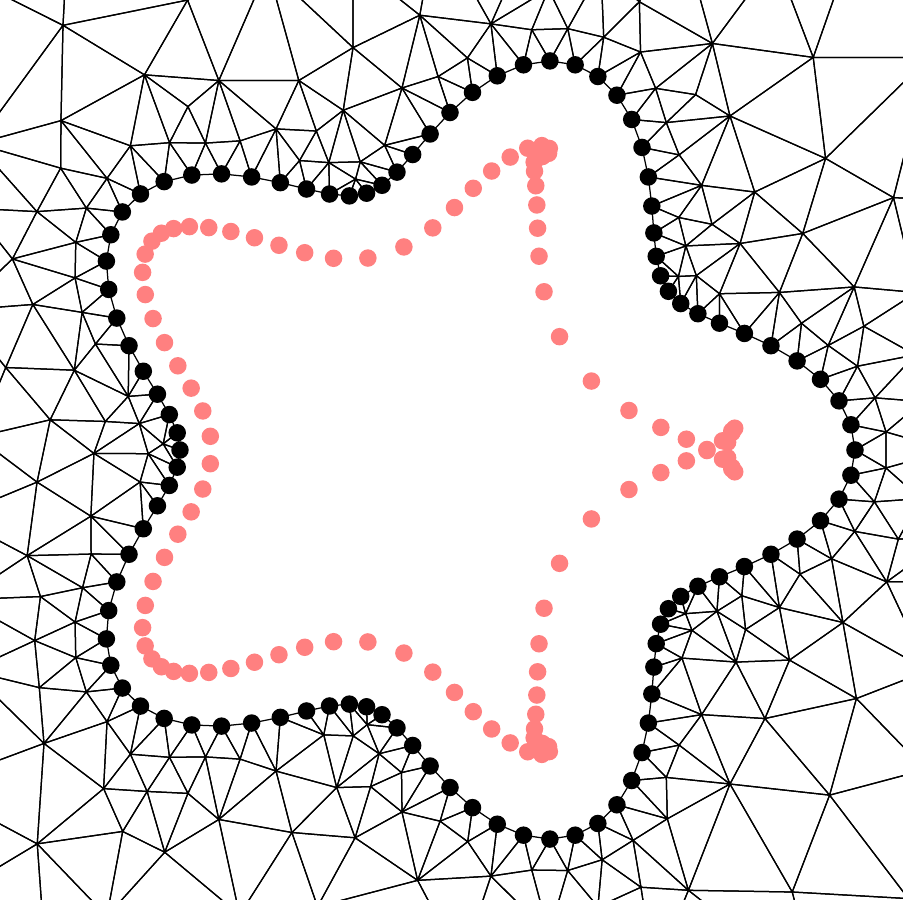}
            \hspace{3mm}
            \includegraphics[width=0.3\textwidth]{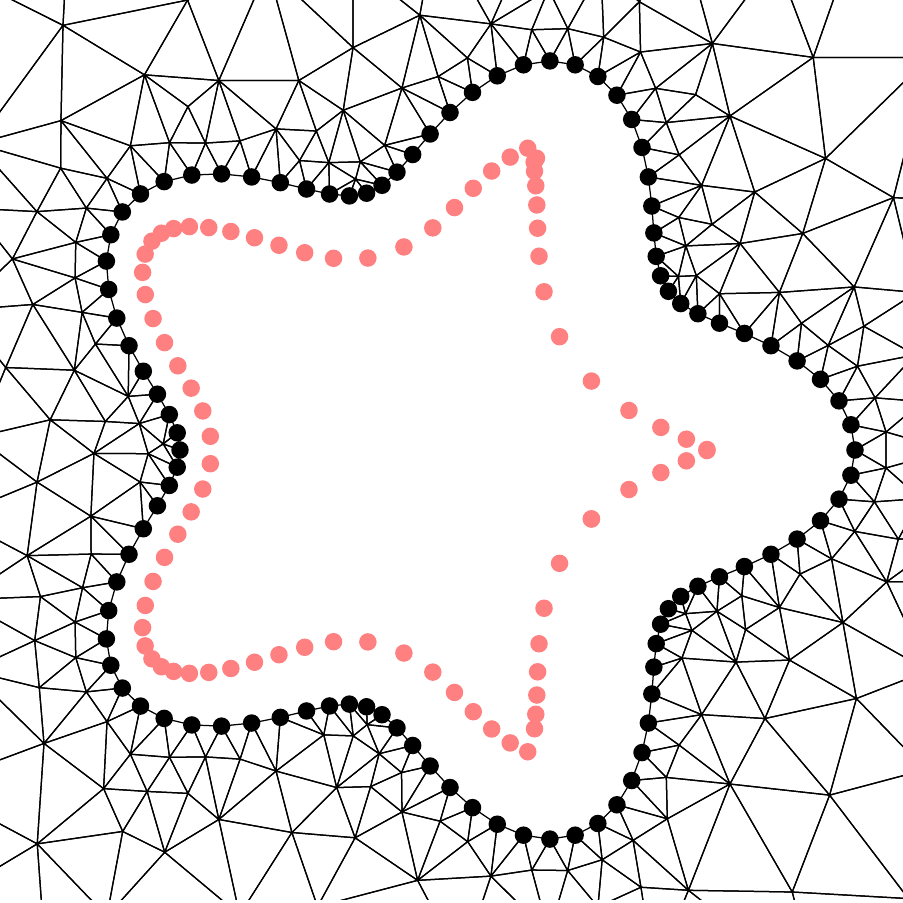}
            % \captionof{figure}{cc}
            
        \vspace{5mm}
        \end{minipage}
\end{enumerate}

\section{Results}

\begin{figure}
    \centering
    \includegraphics[width=\textwidth]{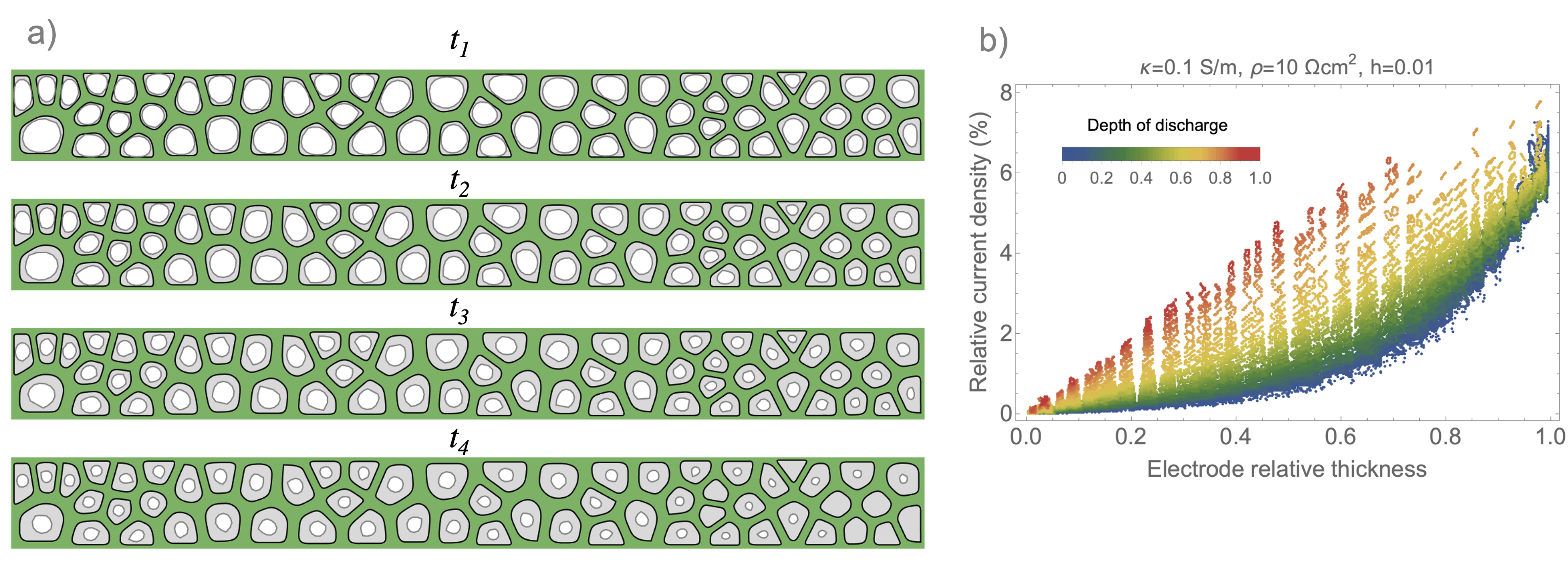}
    \caption{a) A representative 2D slice of a porous anode, shown in green, is filled with lithium (gray region) during battery discharge. The porous structure is made of a mixed electron/Li-ion conducting material with the following properties bulk Li-ion conductivity $\kappa = 1$~mS/m and surface impedance $\rho = 10 \Omega$~cm$^2$. 
    The model is able to track the electro-deposition front inside the pores and predict the evolution in current density distribution over time.
    b) Distribution of the relative current density across the electrode thickness. The color indicates the depth of discharge (DOD) according to the legend. The maximum value is achieved at the separator interface, where the horizontal axis indicates a relative electrode thickness of 1. The current decays away from the separator. The trend changes from exponential to linear around DOD=0.5. }
    \label{fig:results_1}
\end{figure}

\begin{figure}
    \centering
    \includegraphics[width=0.75\textwidth]{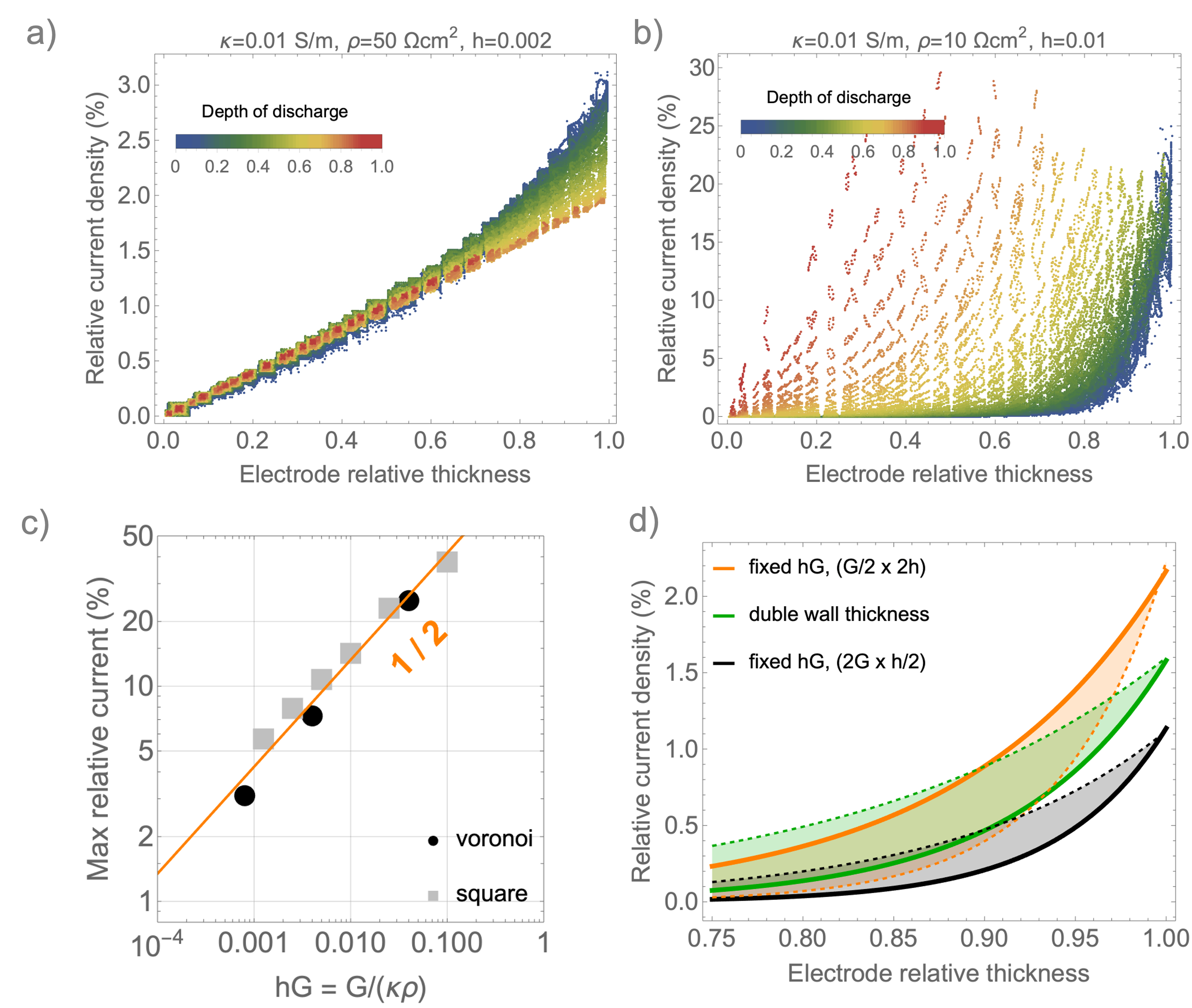}
    \caption{a, b) The current density distribution is plotted, as a fraction of the applied value in the galvanostatic test. The horizontal axis indicates the relative electrode thickness, ranging from zero (current collector) to one (where the separator is located). The effect of bulk and surface conductivity of the porous microstructure is highlighted by the difference in the two plots (the values of $\kappa$ and $\rho$ are listed on each figure). Low bulk conductivity tends to concentrate the current close to the separator interface with lithium progressively filling a few pores at the time (figure~b). Conversely figure~a shows a linear distribution of the current, with a larger surface area engaged in electro-deposition and a lower peak current.
    c) Scaling of the peak current with the dimensionless quantity $hG$ spanning over several order of magnitudes. The convective parameters $h$ is the ratio between surface and bulk conductivity and it appears in the boundary condition Eq.~\ref{eq:robin}. $G$ is the average pore size. In addition to the Voronoi structure shown in Fig.~\ref{fig:results_1} a square lattice morphology was analyzed, and the results are marked by black circles and gray squares respectively. In both cases the peak current shows a clear scaling with respect to the dimensionless parameter as $hG^{1/2}$.
    d) The images shows the sensitivity of the current distribution away from the separator with respect to the wall thickness and the specific values of $h$ and $G$, assuming that the product $h \cdot G$ is constant. The curves refer to DOD=0, and the baseline case is marked by a solid line. Comparing solid the dashed lines of the same color shows that doubling the wall thickness and decreasing the pore size tend to make the current more uniform and less concentrated at the separator interface.
    }
    \label{fig:peakCurr_scaling}
\end{figure}

In Fig.~\ref{fig:results_1}~a, the sequence of four images illustrates how the deposition front moves over time, with Li-metal progressively filling the cells of the porous electrode structure. The cells closer to the separator (on the right) are filled first. The current distribution depends on the bulk conductivity of the foam $\kappa$ and on the impedance at the Li/vapor charge-transfer interface $\rho$. 
For this example the following values were chosen $\kappa = 0.1$~S/m and $\rho = 10 \Omega$~cm$^2$~\cite{doi:10.1021/acsaem.9b02101}. 
The relative current distribution through the electrode thickness is illustrated in Fig.~\ref{fig:results_1}~b, as a function of the depth of discharge (DOD). (At the beginning of lithiation, the state of charge is 1 and the DOD is 0.)
Initially, the local deposition current density reaches the maximum value of $\approx 7.2\%$ of the applied current, and it decays exponentially away from the separator (blue points). This is due to the increase in surface area provided by the porous electrode structure.
Around DOD=0.5, a steady state regime is reached with a linear distribution of the current within the electrode thickness (yellow points in Fig.~\ref{fig:results_1}~b).
The time for establishing the linear regime depends on the conductivity and interfacial impedance.
For instance, Fig.~\ref{fig:peakCurr_scaling}~a shows that a 5-fold increase in interfacial impedance (from 10 to $50 \Omega$~cm$^2$) accelerates the transition to the linear regime.
A moderately larger overpotential at the interface has the effect of homogenizing
the current. However, high interfacial impedance is detrimental to the battery efficiency.

Ideally if the current is uniformly distributed throughout the microstructure, the maximum value drops because the entire interface is equally active for Li-deposition. 
Decreasing the value of Li-ion bulk conductivity $\kappa$ leads to higher local currents and tend to concentrate  Li-deposition to a few cells at the time (Fig.~\ref{fig:peakCurr_scaling}~b). 

The maximum relative current typically occurs at the beginning of lithiation and it serves as a first metric to evaluate the porous electrode design. 
To guide the design of resilient high-energy density systems, we need to identify what combination of material and structural properties prevents current localization at the separator interface, as this can have a negative impact on durability.
To answer this question, we condense the structural and material properties into the dimensionless quantity $hG$, i.e., the product between the average grain size $G$ and the convective term $h$ (having units of 1/length).

The plot in Fig.~\ref{fig:peakCurr_scaling}~c illustrates the scaling of the peak current density at the beginning of discharge $i_{max}$ as a function of the dimensionless parameter $hG$. 
For Voronoi and square lattices (black and gray markers, respectively), we found that the maximum relative current scales approximately as $(hG)^{1/2}$.
Only a small difference due to the specific cell geometry was observed. The two microstructures small differences in tortuosity and wall thickness. The square lattice has tortuosity 1, because the walls are straight and orthogonal to the separator. 

The wall thickness does not affect the peak current at the separator, but it modifies the current decay away from the separator, as shown in Fig.~\ref{fig:peakCurr_scaling}~d (green solid and dashed line). Doubling the wall thickness favors a more homogeneous current density.  
The curves in Fig.~\ref{fig:peakCurr_scaling}~d
refer to the conditions at the beginning of discharge.
Over time, the electrode walls and previously-deposited lithium are both a pathway for Li-ion conduction, so the wall thickness has strongest impact at the beginning of lithiation.
Changing the values of pore size $G$ and of the convective coefficient $h$ in such a way that their product remains constant only affects the trend of the current away from the separator, but not its peak value. The orange and black lines in Fig.~\ref{fig:peakCurr_scaling}~d indicate that decreasing the pore size is generally a good strategy to diffuse plating over a larger area.

Fig.~\ref{fig:peakCurr_scaling}~c suggests that reducing the average cell size and increasing Li-ion conductivity both contribute to lowering $i_{max}$. This result is intuitive, however modeling allows for quantitative metrics in design and material selection.
 This analysis show that the requirement for pore-size distribution depends on the conductivity of the solid-electrolyte material. 
 For highly conductive solid-electrolytes like LLZO ($\kappa \sim 1$~mS/cm), if the average pore diameter is $10 \mu$~m~\cite{doi:10.1021/acsami.9b11780} and interfacial impedance in the order or $10 \Omega$~cm$^2$, the local maximum current density is about 5\% of the global current density.
 This significant reduction in current explains why mesoscale architectures can be effective in improving the cycle life of solid-state batteries with Li-metal, if the solid-electrolyte material is highly conductive.
The results in Fig.~\ref{fig:results_1} and Fig.~\ref{fig:peakCurr_scaling} indicate that the local electrochemical conditions change significantly with the SE properties and it is difficult to engineer the microstructure without an accurate model. 
  
\section{Discussion}

A mixed-conducting foam is investigated as a method  to provide structure and resilience to high-energy density solid-state batteries with a Li-metal anode.
The porous electrode structure greatly expands the surface area of the charge-transfer interface, reducing the local current density to a fraction of the applied planar current.
The anode microstructure provides a conductive scaffold that maintains accessibility of active materials during charge and discharge
With closed cells, the deposition front is confined within the cells, preventing dendrites from puncturing the separator and shorting the cell. If the deposition front becomes unstable dendrites tend to grow inward rather than in the direction of the separator.
Nevertheless, it is currently unknown whether tension will be imposed onto the porous structure as the cell become full of lithium.  This may result in a failure mode that we have not treated here.
If a filled pore causes rupture of the matrix, it would not have an effect on capacity, but
it would increase tortuosity and potentially create a pathway for dendrite growth in the direction of the separator.

Surface irregularities in the boundary layer can be included to study the electrochemical conditions for their amplification, and predict the velocity of dendrite propagation.
The adaptive re-meshing features make this model well equipped to study whether the growing interface is stable or not. These effects are difficult to embed in a commercial finite element software, as they require changes in the topology frequently during the simulation.

\begin{figure}
    \centering
    \includegraphics[width=\textwidth]{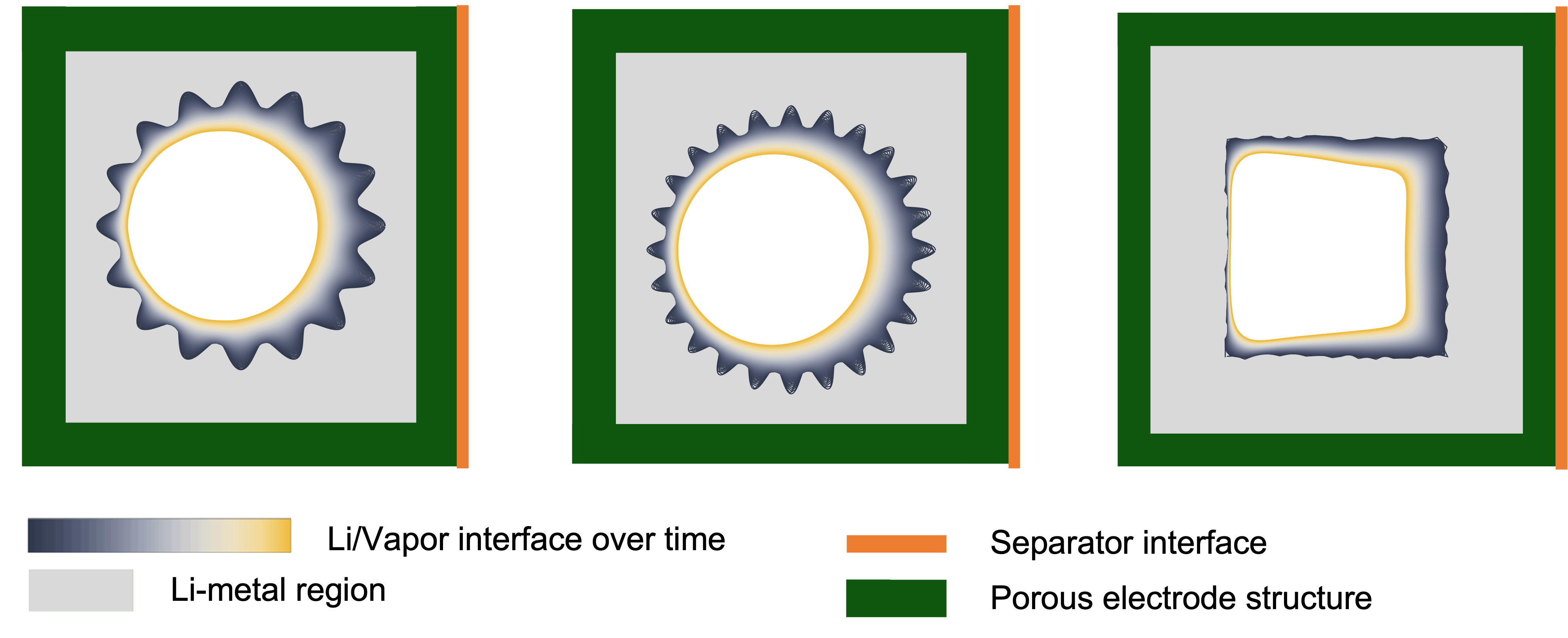}
    \caption{The images show three examples of rough Li/Vapor interface evolving into a smooth boundary. On the left and center images, the initial boundary was chosen to be a sinusoidal wave of different wave length. On the right image, the instability is created by random displacement of the nodes at the Li/Vapor boundary.
    In all the three examples, the analyses indicate that the Li/Vapor interface is stable with respect to non-uniform plating and surface roughness.}
    \label{fig:instability}
\end{figure}

A stability analyses of the Li/Vapor interface was carried out, with some of the results shown in Fig.~\ref{fig:instability}.
In the case of the left and center image a sinusoidal boundary is assumed at the beginning of the simulation. Driven by the local electric field, lithium tends to fill the grooves  because  the current density is higher in the regions of higher curvature. The amplitude of the instability decreases over time, and the interface turns into a circle (marked in yellow in Fig.~\ref{fig:instability}). 
The circle is off-centered with respect to the square lattice because the current is higher towards the separator (marked by the orange boundary in Fig.~\ref{fig:instability}).
Finally, a rough interface was generated by random displacement of the points along a square boundary, as shown on the right image of Fig.~\ref{fig:instability}. Also in this case, the instability decays over time and the Li-/Vapor interface becomes smooth. 

These analyses indicated that Li-dendrite growth can be suppressed by microstructural design. 
However, considerations of mechanical robustness should be taken into account.
Failure mechanisms originated from surface flaws and from the pressure exerted on the surrounding walls as the pores are filled with lithium remain to be investigated.
This requires combining deposition front tracking (presented here) with fracture mechanics, while capturing the constitutive behavior of Li-metal and of scaffold material to accurately predict stress. 
Analyses of the mechanical response will allow to optimize the electrode microstructure for both conductivity and resilience.

% Explicit tracking of the deposition front presents some advantages, as compared to phase field methods, to study the mechanical interaction of multiple dendrites growing from a convex surface and between the interaction with the walls. 
% When two dendrites come into contact they do not simply merge, but they deform exerting pressure on the surrounding walls. 

% The local stress field within each cell tends to self-regulate the local current density and prevent fracture. We calculate that a pressure in the order of 1-5 MPa is sufficient to locally suppress plating.

\section{Conclusions}

The model suggests that a porous electrode matrix may provide means to reduce shorting by confining Li plating into pores.  During discharging, Li plating proceeds inwards to fill the pore.  The simulation suggest that the growth is morphologically stable--it is possible that deleterious effects of dendrite growth might be avoided by microstructural design of the electrode matrix.  

The capacity of the battery is determined by the availability of pore volume.  However, the kinetics of charging are affected by the
tortuosity of the matrix and the surface area to volume ratio of the pores.
Simulation provides scaling approaches to determine what features of a porous electrode are the most important, and how to tune the design to promote uniform current distribution throughout the electrode thickness.

The primary contribution of this paper is the demonstration of an finite element method that can be used and extended to other Li-battery structures~\cite{Delattre_2018, doi:10.1021/acsaem.8b00962, https://doi.org/10.1002/aenm.201802472}.

\section*{Author Contributions}

GB conceived the presented idea and took the lead in writing the manuscript.
GB and WCC developed modeling framework. TS contributed ideas on fabrication requirements and  economic aspects of the battery market. MB advised on the modeling approach. All authors discussed the results and contributed to the final manuscript.

\bibliographystyle{unsrt} 
\bibliography{1_porousElectrode}

\end{document}